\documentclass[conference,10pt]{IEEEtran}

\usepackage{comment}

\usepackage{booktabs}
\usepackage[utf8]{inputenc}
\usepackage[ruled,vlined]{algorithm2e}
\usepackage{microtype}
\usepackage{graphicx}
\usepackage{paralist}
\usepackage{tabularx}
\usepackage{balance}
\usepackage{multicol}
\usepackage{multirow}
\usepackage{pbox}
\usepackage{enumitem}
\usepackage{colortbl}
\usepackage{pifont}
\usepackage{xspace}
\usepackage{url}
\usepackage{tikz}
\usepackage{rotating}
\usepackage{balance}
\usepackage{float}
\usepackage[TABBOTCAP]{subfigure}
\usepackage{tcolorbox}
\usepackage{stackengine}
\usepackage{paralist}
\usepackage{xcolor}
\usepackage{listings}
\usepackage{ifthen}
\usepackage{amssymb}

\newcommand{\ie}{\emph{i.e.,}\xspace}
\newcommand{\eg}{\emph{e.g.,}\xspace}

\newboolean{showcomments}
\setboolean{showcomments}{true}

\ifthenelse{\boolean{showcomments}}
  {\newcommand{\nb}[2]{
    \fbox{\bfseries\sffamily\scriptsize#1}
    {\sf\small$\blacktriangleright$\textit{#2}$\blacktriangleleft$}
   }
  }
  {\newcommand{\nb}[2]{}
  }

\begin{document}


\title{Studying, Identifying, and Fixing Hidden Technical Debt in AI-Intensive Cyber-Physical Systems}


\author{
\IEEEauthorblockN{Beena}
\IEEEauthorblockA{\textit{University of Sannio}\\
Benevento, Italy \\
b.rai@studenti.unisannio.it}
}
\maketitle

\thispagestyle{plain}

\begin{abstract}
Artificial Intelligence (AI) components are increasingly pervasive in several software systems, including Cyber-Physical Systems (CPSs). AI-CPS are used in several domains, including autonomous vehicles, industry, home automation, robotics, and healthcare. Being composed of hardware, AI components, and conventional modules, AI-CPS can exhibit technical debt (TD) that is peculiar and potentially more challenging than that of conventional systems.
This thesis aims to characterize AI-CPS TD and propose approaches for its identification and repair. 
In a first phase, we characterize AI-CPS TD by analyzing AI ecosystems and AI-CPS repositories, as well as interviewing developers. Based on the acquired knowledge, we define approaches to identify and mitigate such TD. Finally, we plan to develop and validate an automated tool that supports agentic AI solutions to monitor, govern, and repay AI-CPS TD.
\end{abstract}





\section{Introduction}
Nowadays, several kinds of software systems are governed by Artificial Intelligence (AI)-intensive components. Among others, such systems include Cyber-Physical Systems (CPSs), \ie systems composed of both software and hardware/mechanical components. Given the complex interactions among software, firmware, and hardware/mechanical components, previous research has shown that CPSs require specialized development processes \cite{TorngrenS18}. This includes, for example, the use of digital twins, tailored continuous integration and delivery (CI/CD) \cite{ZampettiTPPCP23}, and testing \cite{10976605,SomersDWWH23}. 

Modern CPSs also include AI components. Examples of AI-intensive CPSs (AI-CPSs) include autonomous vehicles, industrial or home automation systems, robotic systems, and healthcare automation systems. The AI components themselves have been shown to be responsible for specific types of technical debt (TD) \cite{pepe2024taxonomy}. We conjecture that the interaction between CPS components and AI components may lead to specific challenges, for example, regarding how the AI autonomously enacts a CPS hardware/mechanical component, the extent to which this interaction accounts for real-time hardware requirements, includes the presence of suitable safeguards, and, last but not least, accounts for legal/regulatory requirements.

This PhD project aims to address this gap by answering the following question:

\begin{quote}
\textit{\textbf{What kind of specific technical debt occurs in the development of AI-CPSs, and how can it be automatically repaid?}}
\end{quote}

The PhD work is part of the \emph{InnoGuard: ``Hybrid and Generative Intelligence for Trustworthy Autonomous Cyber-Physical Systems"} Marie Skłodowska-Curie Doctoral Network\footnote{\url{https://innoguard.eu/}}, and it is currently at the beginning of its second year. The research will follow a multi-stage approach.  The project begins with mining Hugging Face and GitHub repositories and a practitioner interview study on the real-world use of AI pre-trained components in software (particularly CPS) automation. The outcome of this first phase will be a taxonomy characterizing AI-CPS TD, developers' challenges in AI-CPS development, and current TD-repairing strategies. The empirical characterization serves as a foundation for automated TD classification and mitigation recommendations. Lastly, we plan to develop agentic AI-based governance for AI-CPS TD continuous identification and classification, monitoring, effect estimation, and repair.

The predicted outcome of this research is, first, a systematic understanding of TD in AI-CPSs, its introduction, root causes, and current mitigation strategies. In a second stage, the thesis will provide AI-CPS developers with a set of approaches for automated TD management, from detection to repayment. Moreover, as indicated by previous work, CPS development---and AI-CPS development---may require suitable and diverse development skills, the thesis' outcome will aid ``democratizing" AI-CPS development and maintenance.


\section{Background And Related Work}
``Technical debt" (TD) refers to the long-term maintenance cost initiated by short-term design, implementation, or performance decisions in software systems \cite{brown2010managing, kruchten2012technical, tom2013exploration}. Earlier research often focused on source code TD, but subsequent studies have shown that TD can also affect architecture, testing, documentation, and maintenance artifacts \cite{besker2018managing, rios2018tertiary}. A widely studied form of TD is self-admitted TD (SATD), where developers explicitly acknowledge suboptimal solutions through comments or other software artifacts. Prior work has proposed several techniques for detecting and classifying SATD using natural language processing and machine learning \cite{maldonado2015detecting, sierra2019survey}. 

In AI-intensive systems, TD becomes broader and more difficult to localize. Besides source code, debt may arise from other artifacts on external platforms or frameworks \cite{pepe2024taxonomy,sculley2015hidden,tang2021empirical}. 
On top of that, AI-CPSs introduce additional concerns related to the interaction between software intelligence and the physical world, such as \textit{sensor robustness, safety constraints, simulation fidelity, calibration, middleware workarounds, and runtime behavior} \cite{bogner2021characterizing,recupito2024technical}. These forms of debt are not yet fully captured by existing taxonomies.


Some research proposed approaches for TD classification, including recent LLM-based approaches \cite{albuquerque2022managing, li2023automatic}, with less focus on mitigation and governance.
Some other work has investigated how TD can be automatically ``repaid".
Specifically, previous approaches used either deep learning techniques  \cite{ALHEFDHI2024107376}, fine-tuned pretrained models \cite{mastro2023}, or, more recently, LLMs \cite{Sheikhaei2026} to repay TD  in conventional software.


While prior work provides strong foundations for TD detection, classification, and automated repayment, we conjecture that AI-CPSs warrant a distinct characterization and approach. In AI-CPSs, TD emerges from the complex interactions among hardware/mechanical components, firmware, conventional software, and AI models. This leads to new TD categories, but also requires TD repayment across heterogeneous artifact types.

\begin{figure}[t]
\centering
\includegraphics[width=0.4\textwidth,height=7cm]{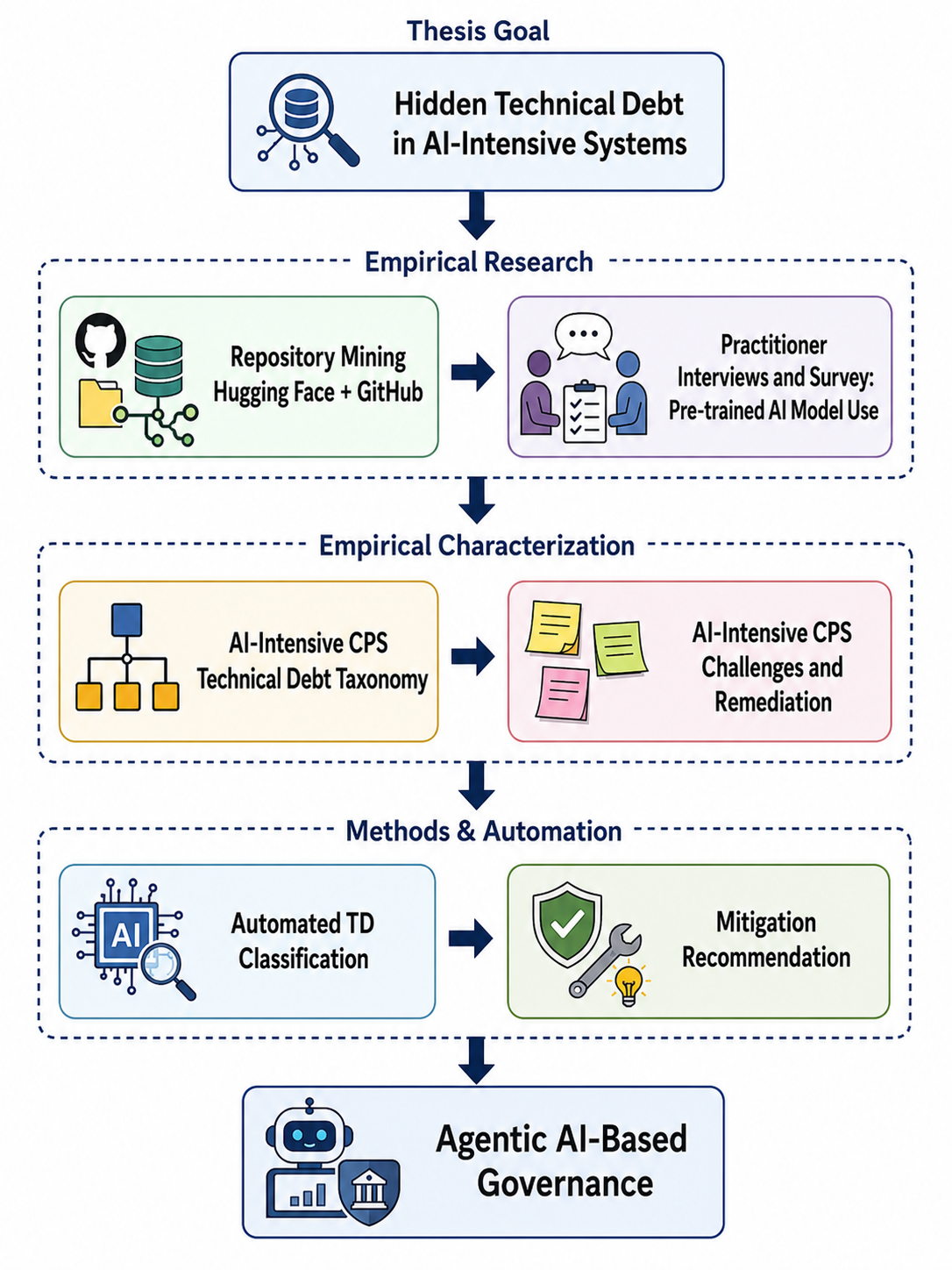}
\caption{Overview of the proposed PhD research approach}
\label{fig:proposal}
\end{figure} 

\section{Methodology}

We first provide an overview of the PhD work methodology. Then, for each step, we detail the proposed methodological or technical solutions.

\subsection{Methodology Overview}
The methodology overview is depicted in Fig. \ref{fig:proposal}.
The research consists of three phases. 
In a first phase, we leverage a mixed-method approach to gather empirical knowledge about AI-CPS TD. We mine Hugging Face and GitHub repositories to explore (i) how pre-trained models are chosen, integrated, grown, and maintained, and (ii) the types of TD that developers self-admit in AI-CPSs. Also, we interview and survey developers to understand their adoption and integration processes for ML models, in particular open-weight models, in software systems, especially in AI-CPSs.
This will allow us to produce a taxonomy of AI-CPS-specific TD and identify challenges in AI model development and integration in CPSs.

In a second phase, based on the output of the first phase, we will develop methods for identifying hidden TD and devise strategies for mitigating each type of debt.

The final phase focuses on agentic AI-based governance. This stage analyzes how the classification and mitigation elements can be integrated into an autonomous governance framework for continuous TD management and repayment.


\subsection{Proposed Solutions}



\subsubsection{Phase 1 - Repository Mining  of AI Model Ecosystems}
The first phase of the project is to gather empirical knowledge about AI-CPS-specific TD. 
First, we perform an empirical repository mining of AI model ecosystems and AI-CPS repositories. On the one hand, we are studying the AI model development process by collecting and analyzing information from Hugging Face models and their linked GitHub repositories. The study evaluates repository links, commits, contributors, file changes, documentation updates, metadata changes, issues, and commit synchronization. 
On the other hand, we are mining AI-CPS repositories to identify and characterize AI-CPS-specific TD, distinguishing AI-CPS-specific instances from those related to conventional and ML-specific TD. Examples of AI-CPS-specific TD concern the relations between software intelligence and the physical world, i.e., \textit{sensing, actuation, motion control, planning, calibration, localization, trajectory generation, physical modeling, safety constraints, robot-environment interaction, simulation fidelity, coordinate transformations, and CPS-specific runtime behavior}.


To complement the mining studies and to understand in depth the experiences that are not directly visible in repository data, we are conducting interview-based studies with practitioners. We investigate how pre-trained AI models are selected, integrated, adapted, deployed, and maintained in real-world software systems from the perspectives of engineers and researchers, and how they work with them. We also investigate the challenges they face in developing AI-intensive systems, particularly AI-CPS.

The first outcome of Phase 1 will be a taxonomy of AI-CPS-specific TD categories, complementing those previously developed for conventional systems \cite{bavota2016large} and ML systems~\cite{pepe2024taxonomy}.


Alongside the taxonomy, we will compile a catalog of the main challenges developers encounter in AI-CPS development, along with possible mitigations and TD repayment strategies. 




\subsubsection{Phase 2 - Methods for Automated Classification and Empirically Grounded Mitigation}

The second phase develops methods to automate identification, classification, and mitigation of ``hidden" AI-CPS TD, \ie TD that has not been admitted by developers. Concerning the TD identification and classification, we develop an agentic-based approach that takes as input the AI-CPS taxonomy and the taxonomies of conventional and ML-specific TD definitions. The approach will leverage metrics and static analysis warnings extracted from the code, alongside the code itself, as well as hardware-related information mined from the repository and ML model information. For the latter, the approach will extract and categorize information from model cards, building on prior literature on model card analysis~\cite{BhatCHLNZKG23}.

The elicited AI-CPS TD taxonomy will serve as the main reference framework. 
More specifically, the identification approach will analyze multiple sources of repository evidence rather than relying solely on explicit TD comments, including CPS-specific evidence such as sensor and actuator descriptions,  middleware settings, simulation files, calibration parameters, safety constraints, deployment scripts, and resource-related information. From the ML side, it will analyze model cards, training and inference scripts, datasets, preprocessing pipelines, and hyperparameters. By combining these sources, we will identify indicators of hidden TD across software, AI/ML, and CPS components.


Mitigation strategies will be empirically identified from the two phases above. We will analyze remediation strategies to observe how developers address debt-related problems in repositories. We will also use practitioner feedback to understand how developers handle challenges in model selection, integration, compatibility, documentation, deployment, reliability, and maintenance. These sources will be used to map debt categories to possible challenges and remediation actions.

\subsubsection{Phase 3 - Agentic AI-based TD governance}

The last phase of the PhD work integrates the outputs of the previous phases into an agentic AI-based governance approach for continuous TD management in AI-CPSs. The purpose is not only to detect TD once, but also to support constant monitoring, classification, impact assessment, and mitigation recommendations as AI-intensive repositories evolve.

\begin{figure}[t]
\centering
\includegraphics[width=0.4\textwidth,height=5cm]{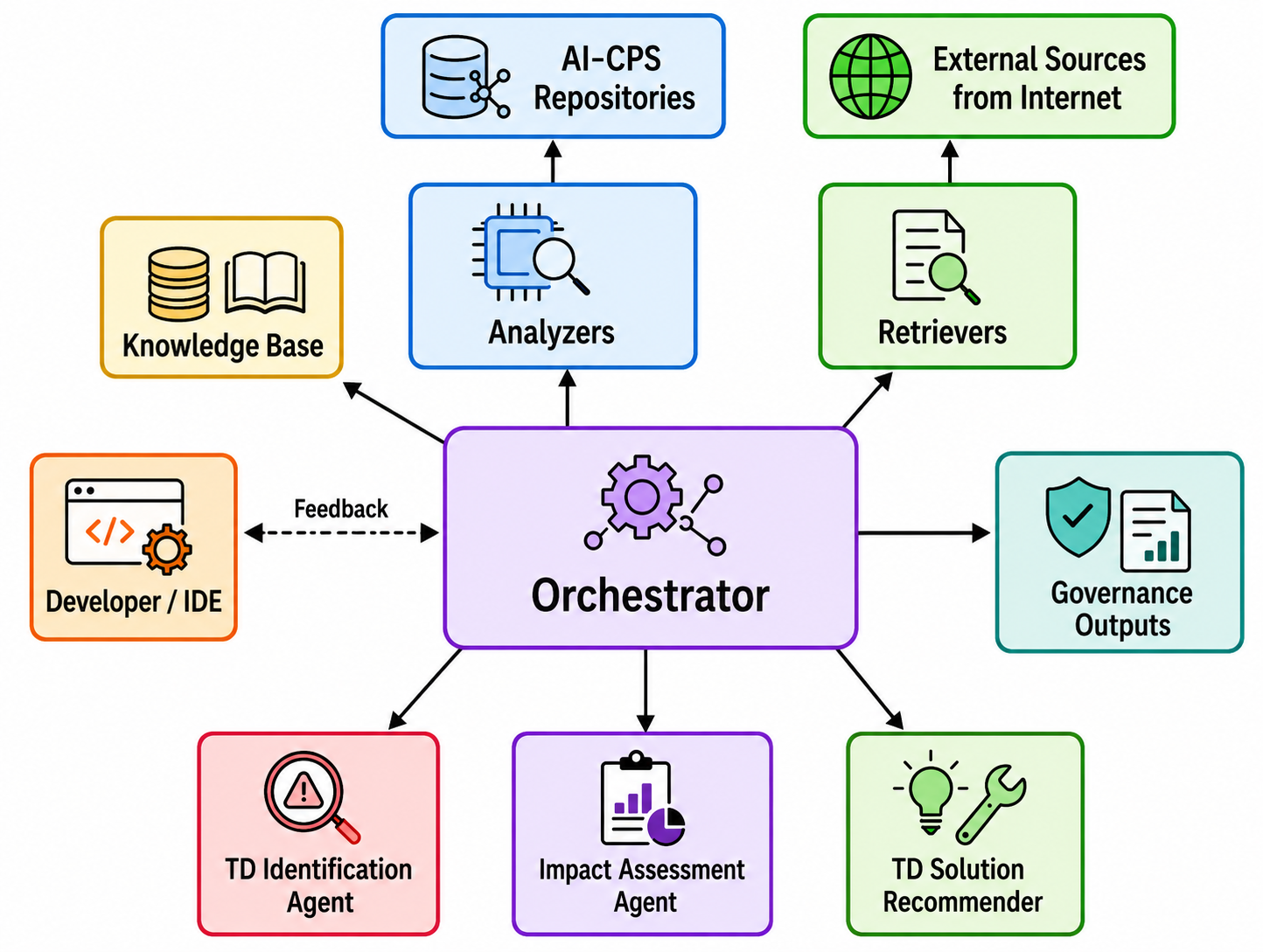}
\caption{Overview of the Agentic AI-based system}
\label{fig:agentic}
\end{figure} 

As shown in Fig. \ref{fig:agentic},
the agentic AI system will be composed of:
\begin{itemize}
 
\item A \textbf{TD identification agent}, which identifies the hidden TD and classifies it along the taxonomy;
\item A \textbf{TD solution recommender} that proposes a remediation solution and, wherever possible, is able to enact it. 
\item A \textbf{challenge and impact assessment} agent will maintain information on the potential effects of the TD in terms of maintainability, reliability, deployability, and other CPS-specific effects. 
\item A \textbf{Knowledge base of empirically acquired evidence}, consisting of the AI-CPS TD taxonomy, developers' challenges, and adopted TD remediation solutions.
\item Specialized  \textbf{analyzers} that mine information from the AI-CPS repositories, and, specifically, analyzers for AI models, firmware, hardware specifications/configurations, and AI-intensive source code.
\item \textbf{Retrievers} that mine data from the Internet, \eg from model/data hubs such as Hugging Face or Kaggle, from software repositories, and from hardware data sheets.
\item An \textbf{orchestrator}, which will handle the entire workflow: monitoring repository changes, invoking analyzers, assigning tasks to specialized agents, verifying output completeness, and producing a final governance report.
\end{itemize}

Notably, the agentic AI solution will be semi-automated, \ie able to interact with the developer through the IDE.
Human feedback will be used to validate classifications and remediation suggestions. This feedback can later be used to refine the taxonomy, improve prompts, update remediation strategies, and make the governance process more reliable.





We plan to empirically validate the agentic AI solution initially through smaller studies in academic settings, \eg by letting students use it in their projects. Then, we will evaluate the Innoguard projects' demonstrators, which involve a humanoid robot and a Leo Rover. This scenario provides a realistic AI-CPS setting that includes sensing, actuation, robot coordination, navigation, runtime monitoring, safety constraints, middleware, and hardware assumptions. It will allow us to assess whether the agentic governance approach can identify AI-CPS-specific debt across software, models, configurations, simulations, and hardware-related artifacts, and whether its impact assessments and remediation suggestions are useful to developers involved.

\section{Expected Contributions}

The PhD work research aims to grow the understanding and management of TD in AI-intensive systems through the following contributions:
\begin{itemize}
    \item \textbf{An empirical characterization of AI model ecosystem evolution and adoption}, resulting from the analysis of ML model GitHub and Hugging Face repositories and from the interview study.
     \item \textbf{A taxonomy of TD in AI-CPS}: The PhD will produce a taxonomy of AI-CPS-specific TD, complementing those for conventional and ML-intensive systems.  
    \item \textbf{Practitioners' perspectives and challenges} on the integration of AI models in software systems and, in particular, in AI-CPS.
   
    \item \textbf{Automated techniques for AI-CPS TD identification, classification, and remediation}: While leveraging knowledge acquired in identifying and repaying conventional TD, the proposed techniques will bring specific identification strategies that look (and fix) into diverse artifacts beyond code, and account for the  AI-CPS specific TD defined in our taxonomy.

    \item \textbf{An agentic AI-based AI-CPS TD governance framework}: The framework will use agentic AI to support repository monitoring, TD identification, classification, impact assessment, and the generation of mitigation recommendations.
\end{itemize}
.

\section{Current status and dissemination plan}
Although the PhD work is in its early stages, two initial results have already been obtained.

\textbf{Empirical study of Hugging Face and GitHub.}at
We completed an empirical study on the development and evolution of ML models, which looked at their Hugging Face and GitHub repositories. Our results show that fewer than 15\% of the most-downloaded Hugging Face models have a development repository on GitHub. We also observed that model cards typically undergo only minor updates, limiting their role as live documentation for ML models. 

\textbf{Initial AI-CPS TD taxonomy construction.}
By analyzing SATD in repositories of 15 open-source 
AI-CPSs, we started creating a taxonomy of AI-CPS-specific TD. This includes, for example, sensor robustness, simulation simplification, safety, middleware or framework workaround, and model integration TD. We plan to submit this work to a software engineering conference in the early fall.

\textbf{Interviews with practitioners} are halfway through being conducted, and we plan to submit an article to a software engineering conference by the fall.

For the forthcoming research, the dissemination will mainly target software engineering conferences and journals, next year for the TD classification and remediation, and during the final year (a journal) for the agentic AI infrastructure.

\section{Acknowledgments}
This work is supported by the InnoGuard Marie Skłodowska-Curie Doctoral Network (Grant Agreement No. 101169233)

\balance

\bibliographystyle{IEEEtranS}

\bibliography{bib.bib}

@inproceedings{BhatCHLNZKG23,
	author       = {Avinash Bhat and
	Austin Coursey and
	Grace Hu and
	Sixian Li and
	Nadia Nahar and
	Shurui Zhou and
	Christian K{\"{a}}stner and
	Jin L. C. Guo},
	title        = {Aspirations and Practice of {ML} Model Documentation: Moving the Needle
	with Nudging and Traceability},
	booktitle    = {Proceedings of the 2023 {CHI} Conference on Human Factors in Computing Systems, {CHI} 2023, Hamburg, Germany, April 23-28, 2023},
	pages        = {749:1--749:17},
	publisher    = {{ACM}},
	year         = {2023}
}

@ARTICLE{10976605,
  author={Mandrioli, Claudio and Shin, Seung Yeob and Bianculli, Domenico and Briand, Lionel},
  journal={IEEE Transactions on Software Engineering}, 
  title={Testing CPS With Design Assumptions-Based Metamorphic Relations and Genetic Programming}, 
  year={2025},
  volume={51},
  number={6},
  pages={1666-1684},
  doi={10.1109/TSE.2025.3563121}}

@article{ZampettiTPPCP23,
  author       = {Fiorella Zampetti and
                  Damian A. Tamburri and
                  Sebastiano Panichella and
                  Annibale Panichella and
                  Gerardo Canfora and
                  Massimiliano {Di Penta}},
  title        = {Continuous Integration and Delivery Practices for Cyber-Physical Systems:
                  An Interview-Based Study},
  journal      = {{ACM} Trans. Softw. Eng. Methodol.},
  volume       = {32},
  number       = {3},
  pages        = {73:1--73:44},
  year         = {2023}
}

@article{SomersDWWH23,
  author       = {Richard J. Somers and
                  James A. Douthwaite and
                  David James Wagg and
                  Neil Walkinshaw and
                  Robert M. Hierons},
  title        = {Digital-twin-based testing for cyber-physical systems: {A} systematic
                  literature review},
  journal      = {Inf. Softw. Technol.},
  volume       = {156},
  pages        = {107145},
  year         = {2023}
}

@inproceedings{TorngrenS18,
  author       = {Martin T{\"{o}}rngren and
                  Ulf Sellgren},
  title        = {Complexity Challenges in Development of Cyber-Physical Systems},
  booktitle    = {Principles of Modeling - Essays Dedicated to Edward A. Lee on the
                  Occasion of His 60th Birthday},
  series       = {Lecture Notes in Computer Science},
  volume       = {10760},
  pages        = {478--503},
  publisher    = {Springer},
  year         = {2018},
  bibsource    = {dblp computer science bibliography, https://dblp.org}
}

@inproceedings{bavota2016large,
  title={A large-scale empirical study on self-admitted technical debt},
  author={Bavota, Gabriele and Russo, Barbara},
  booktitle={Proceedings of the 13th international conference on mining software repositories},
  pages={315--326},
  year={2016}
}

@article{Sheikhaei2026,
author = {Sheikhaei, Mohammad Sadegh and Tian, Yuan and Wang, Shaowei and Xu, Bowen},
title = {A Large-Scale Empirical Evaluation of LLMs for Automated Self-Admitted Technical Debt Repayment},
year = {2026},
publisher = {Association for Computing Machinery},
address = {New York, NY, USA},
issn = {1049-331X},
url = {https://doi.org/10.1145/3796704},
doi = {10.1145/3796704},
note = {Just Accepted},
journal = {ACM Trans. Softw. Eng. Methodol.},
month = feb
}

@article{ALHEFDHI2024107376,
title = {Towards automating self-admitted technical debt repayment},
journal = {Information and Software Technology},
volume = {167},
pages = {107376},
year = {2024},
issn = {0950-5849},
author = {Abdulaziz Alhefdhi and Hoa Khanh Dam and Aditya Ghose},

}

@INPROCEEDINGS{mastro2023,
  author={Mastropaolo, Antonio and Di Penta, Massimiliano and Bavota, Gabriele},
  booktitle={2023 38th IEEE/ACM International Conference on Automated Software Engineering (ASE)}, 
  title={Towards Automatically Addressing Self-Admitted Technical Debt: How Far Are We?}, 
  year={2023},
  volume={},
  number={},
  pages={585-597},
  doi={10.1109/ASE56229.2023.00103}}

@inproceedings{brown2010managing,
  title={Managing technical debt in software-reliant systems},
  author={Brown, Nanette and Cai, Yuanfang and Guo, Yuepu and Kazman, Rick and Kim, Miryung and Kruchten, Philippe and Lim, Erin and MacCormack, Alan and Nord, Robert and Ozkaya, Ipek and others},
  booktitle={Proceedings of the FSE/SDP workshop on Future of software engineering research},
  pages={47--52},
  year={2010}
}

@article{kruchten2012technical,
  title={Technical debt: From metaphor to theory and practice},
  author={Kruchten, Philippe and Nord, Robert L and Ozkaya, Ipek},
  journal={Ieee software},
  volume={29},
  number={6},
  pages={18--21},
  year={2012},
  publisher={IEEE}
}

@article{tom2013exploration,
  title={An exploration of technical debt},
  author={Tom, Edith and Aurum, Ayb{\"u}ke and Vidgen, Richard},
  journal={Journal of Systems and Software},
  volume={86},
  number={6},
  pages={1498--1516},
  year={2013},
  publisher={Elsevier}
}

@article{rios2018tertiary,
  title={A tertiary study on technical debt: Types, management strategies, research trends, and base information for practitioners},
  author={Rios, Nicolli and de Mendon{\c{c}}a Neto, Manoel Gomes and Sp{\'\i}nola, Rodrigo Oliveira},
  journal={Information and Software Technology},
  volume={102},
  pages={117--145},
  year={2018},
  publisher={Elsevier}
}

@article{besker2018managing,
  title={Managing architectural technical debt: A unified model and systematic literature review},
  author={Besker, Terese and Martini, Antonio and Bosch, Jan},
  journal={Journal of Systems and Software},
  volume={135},
  pages={1--16},
  year={2018},
  publisher={Elsevier}
}

@inproceedings{maldonado2015detecting,
  title={Detecting and quantifying different types of self-admitted technical debt},
  author={Maldonado, Everton da S and Shihab, Emad},
  booktitle={2015 IEEE 7Th international workshop on managing technical debt (MTD)},
  pages={9--15},
  year={2015},
  organization={IEEE}
}

@article{sierra2019survey,
  title={A survey of self-admitted technical debt},
  author={Sierra, Giancarlo and Shihab, Emad and Kamei, Yasutaka},
  journal={Journal of Systems and Software},
  volume={152},
  pages={70--82},
  year={2019},
  publisher={Elsevier}
}

@article{sculley2015hidden,
  title={Hidden technical debt in machine learning systems},
  author={Sculley, David and Holt, Gary and Golovin, Daniel and Davydov, Eugene and Phillips, Todd and Ebner, Dietmar and Chaudhary, Vinay and Young, Michael and Crespo, Jean-Francois and Dennison, Dan},
  journal={Advances in neural information processing systems},
  volume={28},
  year={2015}
}

@inproceedings{tang2021empirical,
  title={An empirical study of refactorings and technical debt in machine learning systems},
  author={Tang, Yiming and Khatchadourian, Raffi and Bagherzadeh, Mehdi and Singh, Rhia and Stewart, Ajani and Raja, Anita},
  booktitle={2021 IEEE/ACM 43rd international conference on software engineering (ICSE)},
  pages={238--250},
  year={2021},
  organization={IEEE}
}

@article{recupito2024technical,
  title={Technical debt in ai-enabled systems: On the prevalence, severity, impact, and management strategies for code and architecture},
  author={Recupito, Gilberto and Pecorelli, Fabiano and Catolino, Gemma and Lenarduzzi, Valentina and Taibi, Davide and Di Nucci, Dario and Palomba, Fabio},
  journal={Journal of Systems and Software},
  volume={216},
  pages={112151},
  year={2024},
  publisher={Elsevier}
}

@inproceedings{bogner2021characterizing,
  title={Characterizing technical debt and antipatterns in AI-based systems: A systematic mapping study},
  author={Bogner, Justus and Verdecchia, Roberto and Gerostathopoulos, Ilias},
  booktitle={2021 IEEE/ACM International Conference on Technical Debt (TechDebt)},
  pages={64--73},
  year={2021},
  organization={IEEE}
}

@article{li2023automatic,
  title={Automatic identification of self-admitted technical debt from four different sources},
  author={Li, Yikun and Soliman, Mohamed and Avgeriou, Paris},
  journal={Empirical Software Engineering},
  volume={28},
  number={3},
  pages={65},
  year={2023},
  publisher={Springer}
}

@article{albuquerque2022managing,
  title={Managing technical debt using intelligent techniques-a systematic mapping study},
  author={Albuquerque, Danyllo and Guimar{\~a}es, Everton and Tonin, Graziela and Rodr{\'\i}guezs, Pilar and Perkusich, Mirko and Almeida, Hyggo and Perkusich, Angelo and Chagas, Ferdinandy},
  journal={IEEE Transactions on Software Engineering},
  volume={49},
  number={4},
  pages={2202--2220},
  year={2022},
  publisher={IEEE}
}

@article{polyzotis2018data,
  title={Data lifecycle challenges in production machine learning: a survey},
  author={Polyzotis, Neoklis and Roy, Sudip and Whang, Steven Euijong and Zinkevich, Martin},
  journal={ACM Sigmod Record},
  volume={47},
  number={2},
  pages={17--28},
  year={2018},
  publisher={ACM New York, NY, USA}
}

@inproceedings{pepe2024taxonomy,
  title={A taxonomy of self-admitted technical debt in deep learning systems},
  author={Pepe, Federica and Zampetti, Fiorella and Mastropaolo, Antonio and Bavota, Gabriele and Di Penta, Massimiliano},
  booktitle={2024 IEEE international conference on software maintenance and evolution (ICSME)},
  pages={388--399},
  year={2024},
  organization={IEEE}
}

\end{document}